\documentclass[prl,twocolumn,aps,psfig,epsf]{revtex4}
\usepackage{graphicx}

\begin{document} \title{Spin Screening and
Antiscreening in a Ferromagnet/Superconductor Heterojunction.}
\author{F. S. Bergeret$^{1,2}$ and N. Garc\'{i}a$^{1,2}$} \address{$^1$ Laboratorio de F\'{i}sica
de Sistemas Peque${\it \tilde{n}}$os y Nanotecnolog\'{i}a, CSIC,
Serrano144, E-28006 Madrid\\
$^2$ Laborat\'{o}rio de Filmes e Superf\'{i}cies, UFSC, 88040-900
Florian\'{o}polis, Brazil}

\begin{abstract}
 We present a theoretical study of spin screening  effects in a
ferromagnet/superconductor (F/S) heterojunction. It is shown that
the magnetic moment of the ferromagnet is screened or
antiscreened, depending on the polarization of the electrons at
the Fermi level. If the polarization is determined by the
electrons of the majority (minority) spin band then the magnetic
moment of the ferromagnet is screened (antiscreened) by the
electrons in the superconductor. {  We propose
experiments that may confirm our theory: for ferromagnetic
alloys with certain concentration of Fe or Ni ions there will be
screening or antiscreening respectively.} Different configurations
for the density of states are also discussed.
 \end{abstract}
\maketitle

The coexistence and mutual influence of ferromagnetism and
conventional superconductivity in heterostructures has being
studied intensively in the past years due to the great progress in
preparing high quality multilayered systems (for a review see Ref.
\cite{proshin}). These two long-range  phenomena are antagonistic:
while in the superconducting state electrons form Cooper pairs
with opposite spin, in a ferromagnet the exchange field tries to
aligned the spin of the electrons. Their coexistence in a bulk
material is hardly possible and only takes place for exchange
fields smaller than the characteristic superconducting energy
\cite{lo_state,fulde_state}. The situation changes if the
superconducting and ferromagnetic regions are spatially separated
(e.g. in heterostructures). In this case the coexistence   is
possible even if the exchange field exceeds the value of the
superconducting order parameter $\Delta$, and their mutual
influence is due to the so called proximity effect: when a
superconductor (S) is brought in electrical contact with a normal
metal (N) the superconducting condensate may penetrate into N over
a distance of the order of $\sqrt{D_N/T}$, where $D_N$ is the
diffusive coefficient.  If the normal metal is a ferromagnet (F)
the penetration length is drastically reduced due to the
destructive action of the exchange field $h$ on the Cooper pairs.
Each electron of a pair is in a different spin band. These bands
are shifted by an energy $h$ and therefore if $h$ is very large
the Cooper pair breaks up. In that case the condensate penetrates
into the F region over a distance of the order of $\sqrt{D_F/h}$
and undergoes some characteristic oscillations \cite{radovic2}. In
order to have a stronger proximity effect, i.e. weaker exchange
fields, experimentalists are using dilute magnetic alloys. For
example, in the experiments of Refs. \cite{ryazanov} Cu-Ni alloys
were used in order to observe the change of sign of  the Josephson
critical current in a S/F/S structure. This effect was  predicted
many years ago \cite{bula_kuzi}.

Another interesting effect (the inverse proximity effect) was
studied recently in Ref.\cite{BVE_screening}. The authors proposed
a physical picture according to which some Cooper pairs  share the
electrons between the superconductor and the ferromagnet.  It was
discussed that while the spin of the electron in F prefers to be
parallel to the magnetic moment of F the spin of the electron in S
is automatically antiparallel to the magnetization. In S a
(screening) magnetic moment is induced which penetrates over the
characteristic superconducting length $\xi_s$. Although this
intuitive idea might be true in some cases, it cannot be the whole
story { because the magnetization is not the relevant parameter.
For example for a non itinerant ferromagnet the effect will be
zero or negligible. The reason is that according to the physical
picture the electrons involved in this effect are only those of
the condensate which, as it is well known, are around the Fermi
level (FL). Therefore the screening in the superconductor cannot
be determined by  the magnetization of the ferromagnet which
involves the integral over {\it all} the electrons, but rather by
the polarization of the electrons at the FL as we will show
below.} In Ref.\cite{halterman04} the magnetization of a ballistic
S/F system was studied. However, the authors have not discussed
the inverse proximity effect and instead they found a
magnetization leakage from F to S over distances of the order of
the Fermi wave length. In the present paper we are not interested
in such small scales. The magnetic leakage found in Ref.
\cite{halterman04} can be included by taking a renormalized
thickness of F. Also in Ref. \cite{krivoruchko} leakage of the
magnetic moment into S was reported.

  It is clear
from the physics involved in  F/S junctions that the inverse
proximity effect is  related to the  properties of the conducting
electrons. This implies that the main role is played by the
densities of states (DoS) for electrons with spin up and spin down
at the Fermi level ($\nu_{\pm}(0)$) which in general are
different. The polarization at the Fermi level do  not necessary
have the sign of the magnetization.  In particular the result in
Ref. \cite{BVE_screening} was obtained for the case that the polarization at the
Fermi level is due to majority electrons, and therefore has the
same sign as the magnetization (see Fig. 1). However, it is well
known from band-structure calculations that ferromagnetic metals
show a very complicated band structure and in some cases like Ni,
Co and many other materials, the polarization at Fermi level is
due to minority electrons \cite{bandstructure, chikazumi}.   In
this case at the Fermi level $\nu_-(0)>\nu_+(0)$ (see Fig. 2)  and
therefore according to the physical picture given above the
magnetization induced in the superconductor has the same sign as
in F (antiscreening).

The aim of this letter is to perform a general theory which
explains  this physical picture. We show using the method of the
Green`s function (GFs) that the change of the magnetization of the
system is proportional to the difference of DoS at Fermi level.
The magnetization of the system is reduced if the polarization at
the Fermi level is controlled by spin majority and enhanced if it
is dominated by spin minority. We propose different experiments
and applications which may confirm our predictions. We distinguish
between two types  of ferromagnetic metals : a)  with a conduction
band structure (BS) of type I (Fig. 1) and b) materials with BS of
type II (Fig. 2). The density of states (DoS) at the Fermi level
of the majority spin-band is larger (smaller) than the DoS of the
minority band in the case of materials with BS of type I (II).
\begin{figure}
\includegraphics[scale=0.3]{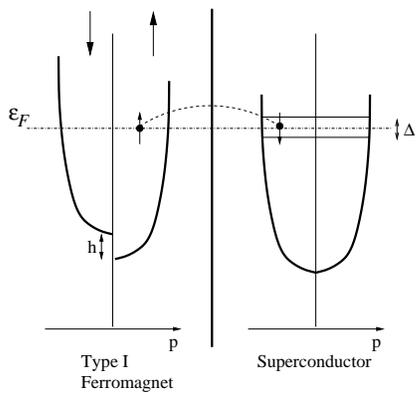}
\caption{S/F system. The ferromagnet shows a type I BS. The DoS at
the Fermi level for majority band (spin up) is larger than the DoS
of the minority band. The two electrons connected by a dashed line
represent a Cooper pair which contributes to the inverse proximity
effect.} \label{Fig.1}
\end{figure}
\begin{figure}
\includegraphics[scale=0.3]{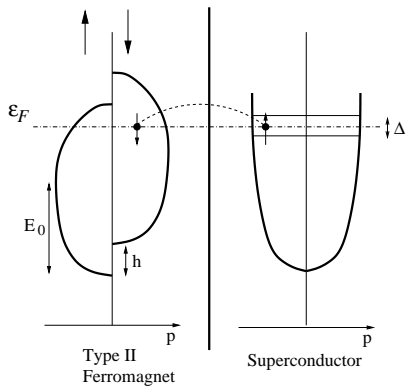}
\caption{S/F system. The ferromagnet shows a type II BS. The DoS
at the Fermi level for minority band  is larger than the DoS of
the majority band.} \label{Fig.2}
\end{figure}
 In order to model both types of materials we chose
a simple model for the ferromagnet which catch the main physics of
the system. We assume that both spin-bands have the same shape and
are shifted by the exchange energy $h$. The Hamiltonian describing
the ferromagnet is given by:
 \begin{equation}
H_F=\sum_{\{p,s,i\}}\left\{ a_{sp}^{+}\left[ \xi _{p}\delta
_{pp^{\prime }}+U_i(p,p')-h(\hat{\sigma}_3)_{ss^{\prime}}\right]
a_{s^{\prime }p^{\prime }}\right\} \label{HamF} \end{equation}
Here $\xi_p$ is the energy of the quasiparticles (counted from the
Fermi energy $\epsilon_F$) and $U_i(p,p')$ is the scattering
potential from the $i$ impurity.  The last term describes the
ferromagnetic interaction which is written in the mean field
approximation and leads to the shift of the spin bands.
 In the free electron model
and defining $E_0$ as the midband energy we assume that the
momentum is
\begin{eqnarray}
 p&=&\sqrt{2mE}\;\;\; {\rm for }\;\; E<E_0\label{dispI}\\
 p&=&\sqrt{2m(2E_0-E)}\;\;\; {\rm for}\;\; E>E_0\label{dispII}
\end{eqnarray}
Of course one can choose another shape for the curves $E(p)$.
However the main results of this paper do not depend on this
choice. Notice also that there may be another type of materials
for which the Fermi energy lies for one spin band above $E_0$ and
for the other spin band below $E_0$.  The generalization of our
results for this case is straightforward.

The Green's functions $G_{\pm}$ for the spin up and spin down
electrons corresponding to the Hamiltonian (\ref{HamF}) are
\begin{equation}
G_\pm(\omega_n,{\bf p})=\left(i\omega_n-\xi_p\mp h/2+({\rm
sgn}\omega/2\tau)\right)^{-1}\; ,
 \end{equation}
where  $\omega_n=\pi T(2n+1)$ is the Matsubara frequency and
$\tau$ is the momentum relaxation time caused by the impurities.
The DoS for spin up and down electrons are
\begin{equation}
\nu_\pm(\omega)=\frac{m}{2\pi^2}\sqrt{2m(\epsilon_F\pm
h+\omega)}\label{nuI}
 \end{equation}
  for energies above $E_0$, and
\begin{equation}
 \nu_\pm(\omega)=\frac{m}{2\pi^2}\sqrt{2m(2E_0-\epsilon_F\mp h-\omega)}
\label{nuII}
 \end{equation}
for energies below $E_0$.  We emphasize that the spin polarization
at the fermi level for materials with BS of type I is positive
while for materials of type II is in the opposite direction.
However, the {\it total} magnetization $M_F$ is obtained by
integration over all $\omega$'s and therefore is positive in both
cases.

The superconductor is described by the usual BCS Hamiltonian in
the mean field approximation
\begin{equation}
\hat{H}_{BCS}=H_0-\sum_{\{p,s\}}\left\{ \Delta a_{\overline{s}\overline{p}%
}^{+}a_{sp}^{+}+c.c.\right\}\; ,  \label{HamS}
\end{equation}
where $H_0$ is the free electron part which contains also
scattering by impurities. $\Delta$ is the superconducting order
parameter. The index $s$ denotes spin and $p$ momentum. The Cooper
pairs forming the condensate have total momentum equals zero and
are in singlet state ($\overline{s}$ and $\overline{p}$ stay for
$-s$ and $-p$ respectively) .
 We are
interested in the inverse proximity effect, in particular how the
magnetization $M$ of the system changes due to the presence of the
superconducting correlations.  The total Hamiltonian of the system
is $H=H_F+H_{BCS}$. Finding the GF for the S/F structure is a
quite difficult task and some simplifications have to be made. We
use here the well known quasiclassical approach (see e.g.
\cite{lo_book}). The quasiclassical Green's functions are obtained
by integrating the microscopic ones over $\xi_p$ and only content
information about electrons close to the Fermi surface. This
restriction does not limit our analysis since only the electrons
in a narrow region around the Fermi level of the order of
$\Delta$, participate in the proximity effect.
 In order to obtain the equations for the quasiclassical
GF one  assumes that all energies involved in the problem are
small in comparison to $\epsilon_F$, in particular
$h\ll\epsilon_F$. The quasiclassical equations are derived in many
papers and therefore we skip here the derivation (see for example
Ref.\cite{lo_book}). Tagirov generalized these equations for the
case that the momenta at the Fermi level $p_{F\pm}$ of both spin
bands are different \cite{tagirov_C}. According to Eqs.
(\ref{nuI}-\ref{nuII}) $p_{F\pm}=2\pi^2\nu_{\pm}(0)/m$.
 For a
diffusive system  one obtains the general Usadel equation
\cite{tagirov_C,usadel}
\begin{equation}
D\nabla \left( \mbox{$\check{g}$}\nabla \check{g}\right) -\omega_n
\left[
\mbox{$\hat{\tau}_3$}\mbox{$\hat{\sigma}_0$},\mbox{$\check{g}$}\right]
+iv_F\delta p_F \left[
\mbox{$\hat{\tau}_3$}\mbox{$\hat{\sigma}_3$},\mbox{$\check{g}$}\right]
=-i\left[ \check{\Delta},\mbox{$\check{g}$}\right] \;.
\label{Usad}
\end{equation}
 The GF $\check{g}=\hat{g}\hat{\tau}_3+\hat{f}i\hat{\tau}_2$ is a
4$\times$4 matrix in the spin ($\hat{\sigma}$) and particle-hole
space ($\hat{\tau}$) and $\omega_n=\pi T(2n+1)$ are the Matsubara
frequencies \cite{BVE_josephson}. In the $S$ region $D=D_{S}$,
$\delta p_F\equiv p_{F+}-p_{F-}=0$, and
 $\check{\Delta}=\Delta i\hat{\tau}_2\hat{\sigma}_3$ (the phase of
$\Delta $ is chosen to be zero). While in the F layer $D=D_{F}$,
$\Delta =0$, $\delta p_F\neq0$ and $v_F$ is the Fermi velocity for
vanishing exchange field. The term proportional to $v_F\delta p_F$
is related to the effective exchange field acting on the electrons
at the Fermi level. In the limit under consideration
($h\ll\epsilon_F$) $h=v_F|\delta p_F|$ \cite{tagirov_C}. Note that
the sign of this term depends on wether F has a BS of type I or
II. Equation (\ref{Usad}) is complemented by proper boundary
conditions \cite{kuprianov,tagirov_C}.

In order to avoid cumbersome calculations we make a further
simplification which does not change the qualitative validity of
our results. We assume that the F and S layers are thinner than
the characteristic length of variation of the GFs. In that case
one can average Eq. (\ref{Usad}) over the thicknesses and define
an effective exchange field $h_{eff}\equiv v_F\delta p_F\left( \nu
_{F}^{I,II}d_{F}\right) /\left( \nu
_{F}^{I,II}d_{F}+\nu_{S}d_{S}\right) $, where $\nu _{F}^{I,II}$
are the corresponding DoS at zero value of the exchange field for
case I and II.
 We also define ${\Delta}_{eff}\equiv {\Delta}\left( \nu _{S}d_{S}\right) /\left( \nu _{S}d_{S}+\nu
_{F}^{I,II}d_{F}\right) $. { Within this approximation and under
the assumption that the S/F interface is perfect Eq.(\ref{Usad})
can be transformed into an algebraic equation for $\check{g}$
complemented by the normalization condition $\check{g}^2=1$. The
solution of this set of equations can be found easily (see e.g.
Refs. \cite{BVE_josephson,golubov_tc}).}

Our aim is to calculate the  magnetization per  unit area induced
in the superconductor \cite{BVE_screening}
\begin{equation}
M_S=-i\mu_B\pi\nu_Sd_ST\sum_{\omega_n}{\rm
Tr}\hat{\sigma}_3\hat{g}\; \label{magn},
\end{equation}
where $\mu_B$ is the Bohr magneton. If the F layer is very thin
the expression for the component of $\hat{g}$ proportional to
$\hat{\sigma}_3$ is
\begin{equation}
g_3=- i
\frac{h_{eff}\Delta_S^2}{(\omega_n^2+\Delta_s^2)^{3/2}}\label{g3}
\end{equation}
Inserting this expression in Eq. (\ref{magn}) we obtain for the
magnetization induced in S  at $T=0$ (per unit area)
\begin{equation}
M_S^{I,II}= {\cal N}\mu_B (v_F\delta
p_F)\frac{\nu_F^{I,II}d_F}{\nu_F^{I,II}d_F+\nu_Sd_S}\nu_S d_S\;
\label{final},
\end{equation}
where  ${\cal N}$ is a positive numerical factor of the order of
unity { and we have transform the sum over the Matsubara
frequencies into an integral. For finite temperatures and
according to Eq. (\ref{g3}) the induced magnetization is a
monotonically decaying function of the temperature which vanishes
when $T=T_C$ as expected}.  It was shown in
Ref.\cite{BVE_screening} that the component $g_3$ of the GF
induced in the superconductor penetrates over the length $\xi_s$.
Thus, if the thickness of the superconductor is larger than the
coherence length $\xi_S$, then Eq. (\ref{final}) can be used for
estimates if one substitutes $d_S$ by $\xi_S$.

Equation (\ref{final}) confirms our intuitive picture given in the
introduction. Depending on the sign of $\delta p_F$ which is
proportional to $\nu_+(0)-\nu_-(0)$ the magnetization induced in S
is antiparallel (case I, $\delta p_F>0$) or parallel (case II,
$\delta p_F<0$ )  to the magnetization in F. From Eq.(\ref{final})
one can see  that the maximum induced magnetic moment in S is
related to the density of electrons at the FL $\nu_s$. This
quantity approximately equal to $\Delta.\nu_S$, i.e corresponds to
10$^{-3}$-10$^{-4}$ Bohr magneton per atom. This is very small
quantity  and therefore will be difficult to observed this effect
with usual magnetic material as Fe or Ni. In order to check these
effects one should try with dilute materials, ferromagnetic
semiconductors \cite{ohno} or in materials with very low
magnetization as for example seems to be the case of graphite and
polymerized fullerenes \cite{fulleren}.

{ By deriving Eq.(\ref{final}) we have assumed that the S/F
transparency is high enough. However, it is known that  in many
experiments the S/F interfaces are not perfect and the
transparency may be very low. In this case the proximity effect is
weak and  one can linearize Eq. (\ref{Usad}). This limit was
considered by the authors of Ref.\cite{BVE_screening} for a F
layer with a BS of type I. In that case the induced magnetization
decreases as $R_b^{-2}$ by increasing the interface resistance
$R_b$. This result is also valid in the case of type II BS. The
main difference is that in the latter case and  according to our
theory the induced magnetization will be parallel to the
magnetization of F and hence the total magnetic moment will
increase. Thus,  high values of $R_b$  will suppressed the inverse
proximity effect in both cases. An increase of the interface
resistance can be due to a formation of an oxide layer between the
metals or band mismatch.

We propose possible experiments that will illuminate the
correctness or not of our theory. For ferromagnetic alloys with,
for example, certain iron concentration, as the systems  VFe/V or
PdFe/V used in Refs. \cite{aarts} and \cite{garifullin}
respectively, there will be a screening effect because in these
alloys the majority electrons at the FL aligned with the
magnetization. However, for the case of ferromagnetic alloys with
Ni ions (e.g. the junction NiCu/Nb used in Ref.\cite{ryazanov})
antiscreening will take place due to the fact that the electrons
of Ni at the FL are dominate by minority electrons.}

 If the widths of the conduction band are very different it is
clear from the physical picture that there is no possibility to
have Cooper pairs sharing their electrons between the ferromagnet
and the superconductor because the momenta matching is very bad.
In that case the proximity effect, i.e. the penetration of Cooper
pairs into the F region, is negligible small. However, one can
imagine the situation depicted in Fig. 3, where the exchange field
in F is so strong that the Fermi momenta for electrons with spin
up and down are very different (this is similar to the situation
of a half metal \cite{schoen_half}). For example,  if the width of
the minority band is similar to the width of the band of the
superconductor then according to our theory the inverse proximity
effect will lead to an enhancement of the total magnetic moment,
since only the electrons of the minority band can be paired with
electrons of S. It can also occur that the majority spin-band
width corresponds to the S band width. In that case we predict a
decrease of the total magnetic moment when $T$ is lowered below
$T_c$. Thus, the effect consider in this paper can be used in
order to study the electronic properties of ferromagnetic
materials at the Fermi level. One can perform an experiment by
measuring the magnetization for temperatures above and below the
superconducting temperature. If by lowering the temperature the
magnetization is enhanced, then it is clear that at the Fermi
level the minority spin-band dominates, and viceversa. The
situation depicted in Fig. 3 may correspond to the case of some
high $T_C$ superconductors which in general have very low Fermi
energies.
\begin{figure}
\includegraphics[scale=0.3]{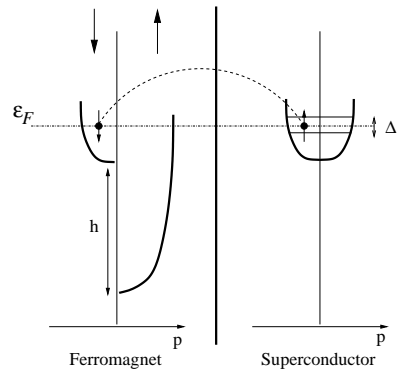}
\caption{S/F structure consisting of a ferromagnet with a large
exchange splitting $h$. The band with of the minority spin-band is
approximately equal to the band width of the superconductor.}
\label{Fig.3}
\end{figure}

One can also use this effect for examining if a ferromagnetic
material  has either its magnetic moments concentrated in  small
regions or  distributed homogenously (see Fig. 4). The number of
magnetic moment which can be screened (or antiscreened) is
proportional to $\Delta \nu_S (\xi_S{\cal S})$, where ${\cal S}$
is the cross section of the magnetized region. It is clear from
our analysis that if the magnetization of F is due to highly
magnetized small regions the relative change of magnetization is
negligible, while if the magnetic moments are homogenously
distributed, the effect of screening or antiscreening might be
more pronounced.
\begin{figure}
\includegraphics[scale=0.3]{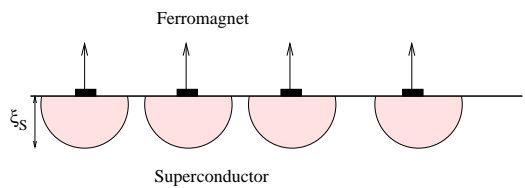}
\caption{S/F systems. The magnetization in F is due to the
magnetic moment of certain regions. The screening or antiscreening
in S is only possible in regions of size $\xi_s$ (circles)}
\label{Fig.4}
\end{figure}

In conclusion we have studied the inverse proximity effect in a
S/F system. Superconducting correlations  leads to the formation
of Cooper pairs which share their electrons between the
superconductor and the ferromagnet. Depending on the polarization
of the electrons at the Fermi level  we predict a screening  or an
antiscreening  of the magnetic moment. If the DoS at the Fermi
level  of the majority band is larger than the DoS of the minority
one then the magnetization of the system is reduced by lowering
the temperature below the superconducting temperature. In the
opposite case we predict an enhancement of the magnetization. Such
effect may be useful to examine the electronic properties at the
Fermi level and the distribution of magnetic moments of
ferromagnetic metals.

{ This work has been supported by the Spanish DGICyT and by the
FP6 EU program. }




\end{document}